\documentclass[a4paper]{jpconf}
\usepackage{graphicx}
\usepackage{amsmath}
\begin{document}
\title{SM antideuteron background to indirect dark matter signals in galactic cosmic rays}

\author{D.M. Gomez Coral$^{1,2}$ and Arturo Menchaca-Rocha$^1$}

\address{$^1$ Instituto de F\'isica, UNAM, A.P. 20-364, CP 10000, CDMX, M\'exico}

\address{$^2$ Department of Physics and Astronomy, University of Hawaii at Manoa, 2505 Correa Rd, Honolulu, HI 96822}

\ead{dgomezco@hawaii.edu.}

\begin{abstract}
Antideuteron production cross-sections estimated using EPOS-LHC with a coalescence afterburner, tuned to reproduce published experimental data over a wide range of energy were used here as input to the galactic propagator code GALPROP, validated with comparing to existing proton, helium fluxes as well as boron-to-carbon ratio data. The resulting near-Earth antideuteron flux, including solar modulation, is compared to previous estimates. An overall factor of two increment in the antideuteron flux is predicted, the origin of which is also discussed. However, this standard model source of antideuteron background still lies well below the AMS-02, and the expected GAPS, sensitivities, as well as the fluxes predicted by several dark matter models.
\end{abstract}

\section{Introduction}
Within the Standard Model (SM), cosmic deuteron formation is understood as the result of nuclear interactions of primary CRs, mainly protons, with the interstellar matter (ISM). This cosmic deuteron source, known as secondary production, is dominated by two contributions: fragmentation of ISM nuclei, in which deuterons are direct products; and the resonant reaction $p+p \rightarrow \pi + d$. Thus, cosmic deuteron flux provides important information about CR propagation in the Galaxy, such as the mean amount of ISM that primary CR’s encounter as they travel from their sources to the Earth. Accelerator experiments allowed the study of a third deuteron production mechanism, explained within the framework of the so-called coalescence model. This applies to free nucleons resulting from CR-ISM interactions, in which residual protons and neutrons lie sufficiently close enough in phase space recombine (coalesce) to form deuterons. Such free nucleons may be the result of p-nuclei fragmentation interactions. But, at sufficiently high energies, p-p and p-nuclei interactions can also create multiple nucleon-antinucleon pairs, generating conditions for the formation of deuterons through the coalescence mechanism. Note that, of the three deuteron-producing mechanisms described, this is the only one that also allows the formation of secondary antideuterons. The result is an antideuteron flux having an energy distribution with a maximum at $\approx 4 $ GeV. Lower energy antideuterons, with maximum at $\approx 1$ GeV and a similarly energy distribution, can also be produced in the interaction of primary protons with secondary antiprotons, with smaller probability. Yet, in that energy range, when all possible mechanisms are taken into account, as illustrated by T. Aramaki et al. \,\cite{Aramaki} and reproduced in Fig.\,1, the predicted abundance of antideuterons, estimated using simplified coalescence model calculations and geometrical cross sections, turns out to be at least one order of magnitude smaller than what several models predict for antideuteron production from dark matter particle-antiparticle annihilation. 

\begin{figure}[h]
\begin{center}
\includegraphics[height=0.3\textheight]{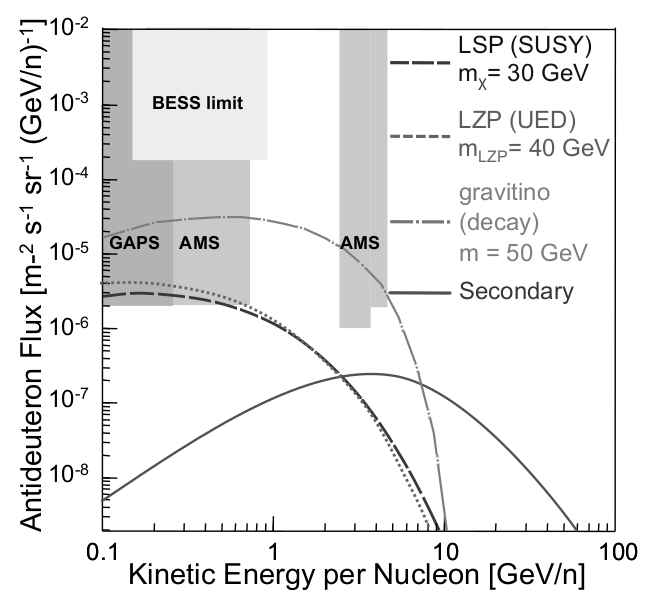}
%includegraphics[height=0.28\textheight]{cap1/dbar_dm_1.png}
\caption{Expected antideuteron flux from three different dark matter candidates\,\cite{Donato1, Baer, Dal} compared to expected secondary production predicted by Ibarra et al. \,\cite{Ibarra} and the corresponding detector sensitivity limits.}
\label{fig1}
\end{center}
\end{figure}

The predicted low energy secondary SM antideuteron-free window ($E_{kin} = 0.1 $ GeV to $1$ GeV) shown in Fig.\,1 generated great interest in dark matter research. Given the continuous improvement of Monte Carlo particle-interaction simulators, as well as the development by our group of an afterburner scheme to include in them the deuteron/antideuteron coalescence mechanism \,\cite{Serradilla}, in Ref. \,\cite{Diego1} we estimated secondary antideuteron cross sections predictions (see Fig.\,2), based on available accelerator data over a wide range of energies.

\begin{figure*}[!htb]
\begin{center}
\begin{tabular}{ll}
\includegraphics[width=7.6cm]{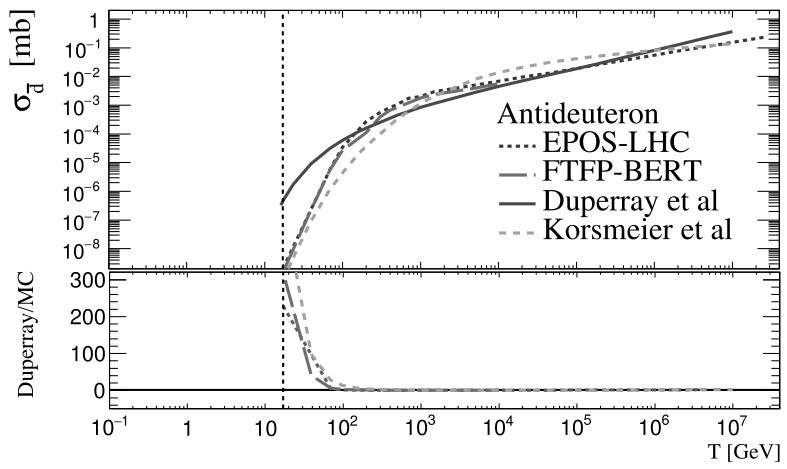}
& \includegraphics[width=7.6cm]{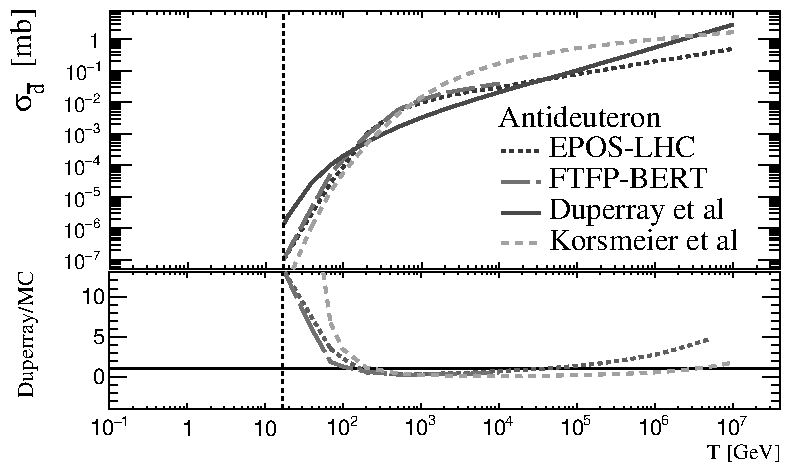}\\
%\includegraphics[width=7.6cm]{cap2/Fig6a.eps} % Eliminar las partes de arriba (deuterones) 
%&\includegraphics[width=7.6cm]{cap2/Fig6b.eps}\\ % Eliminar las partes de arriba (deuterones) 
\end{tabular}
\caption{Antideuteron total production cross-section\,\cite{Diego1} in p+p collisions (left), and in p+He collisions (right). The expected antideuteron cross-section from Duperray's parametrization\,\cite{Duperray} has been added. In the lower panels Duperray and MC predictions for antideuteron are compared. Vertical broken lines represent the antideuteron production threshold.}
\label{fig2}
\end{center}
\end{figure*}

In Fig.\,2, as expected, simplified cross section estimations work best at high energies but show important deviations from our HEP simulators in the low energy region, where the best possible SM background knowledge is required. 

Here, we present preliminary results \,\cite{Diego2} of using the cross sections shown in Fig.\,2 as input to the galactic-transport code GALPROP \,\cite{Strong}, to calculate the near-Earth antideuteron flux. This code incorporates as much realistic astrophysical input as possible, together with the latest theoretical developments to simulate the propagation of cosmic-ray nuclei, antiprotons, electrons, and positrons, and computes diffuse $\gamma$-rays and synchrotron emission in the same framework. 

\section{Transport}
GALPROP\,\cite{Strong} solves the transport equation considering a source distribution and boundary conditions for all cosmic-ray species. The equation includes Galactic wind (convection), diffusive re-acceleration in the interstellar medium, energy losses, nuclear fragmentation, and nuclear decay. The numerical solution of the transport equation is based on a second-order implicit scheme. The spatial boundary conditions assume free particle escape. 
The transport equation is written as

\begin{equation}\label{cp3:eq8}
\begin{aligned}
\frac{\partial f(p,\vec{r},t)}{\partial t}=\vec{\nabla}\cdotp \big(D_{xx}(p, \vec{r})\vec{\nabla}f - \vec{V}f \big) + \frac{\partial}{\partial p} p^2 D_{pp} \frac{\partial}{\partial p} \frac{1}{p^2} f \\
-\frac{\partial}{\partial p}\left[\dot{p}f - \frac{p}{3} (\vec{\nabla}\cdotp\vec{V})f \right] - \frac{1}{\tau_{f}}f - \frac{1}{\tau_{r}}f + q(p,\vec{r},t),
\end{aligned}
\end{equation}

where $f(p;\vec {r}; t)$ is the density of particles in the ISM per unit of total momentum. $D_{xx}$ is the spatial diffusion coefficient, $\vec{V}$ is the convection velocity, $D_{pp}$ is the diffusion coefficient in momentum space describing re-acceleration, $\dot{p}$ is the momentum loss rate, $\tau_{f}$ is the timescale for fragmentation, and $\tau_{r}$ is the timescale for the radioactive decay. The last term of Eq.\,1 represents the source, integrated of all possible secondary sources contributing to a particular isotope species. For antideuterons, the secondary source terms obtained with the corresponding EPOS-LHC cross sections are shown on the left of Fig.\,3. From this, it is concluded the p+p channel represents around $48\%$ of the total flux, while the antiproton channels account for $\approx 5.2\%$ of the antideuteron flux. Note, however, that the latter contribution is dominant below $1$\,GeV. The contribution from He+He has also been included.
On the right side of Fig.\,3, the EPOS-LHC source term ($q_{\bar{\text{d}}}$) is compared to results from previous works\,\cite{Duperray, Ibarra}. These calculations predict a higher antideuteron production in CRs collisions, reaching around twice the value from Ibarra et al.\,\cite{Ibarra} and $\approx 1.2$ times the result from Duperray et al.\,\cite{Duperray}. Furthermore, the maximum of the source distribution is slightly shifted in the antideuteron kinetic energy in comparison to parametrizations, being located at $\approx 5 $GeV/n. This shift is a consequence of competing factors folded into the secondary source term, i.e., the rising production cross-section that dominates at low energies (see Fig.\,2), and the rapidly decreasing incident CR flux that governs the high-energy region. Thus, in contrast with previous calculations, the production cross-section from EPOS-LHC is more significant in the high energy range, causing an increase observed in $q_{\bar{d}}$ and, despite the decreasing incident flux. It also produces a broader distribution with a flattened shape on the top of the curve, and the shifting of its maximum.

\begin{figure}[h]
\begin{center}
\begin{tabular}{ll}
\includegraphics[height=0.29\textheight]{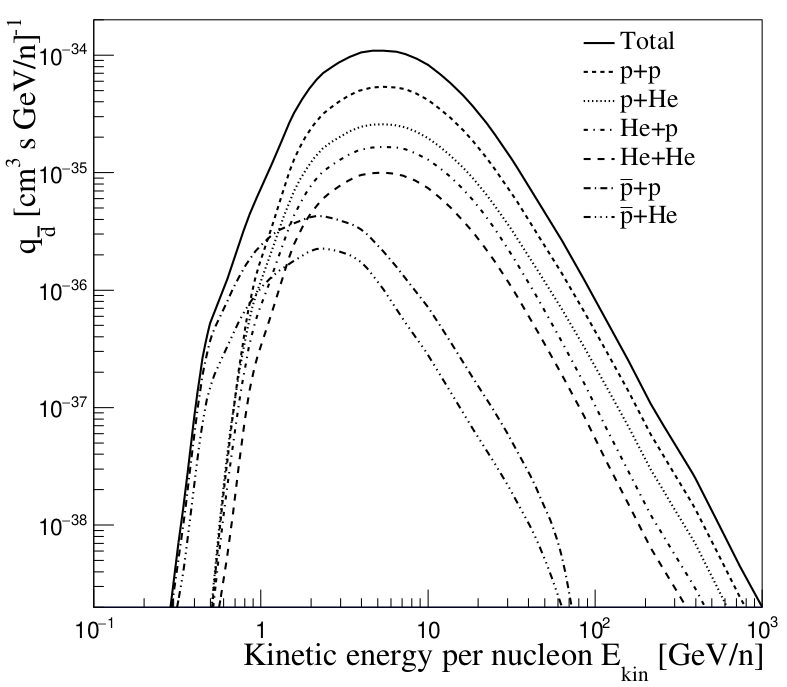}
& \includegraphics[height=0.29\textheight]{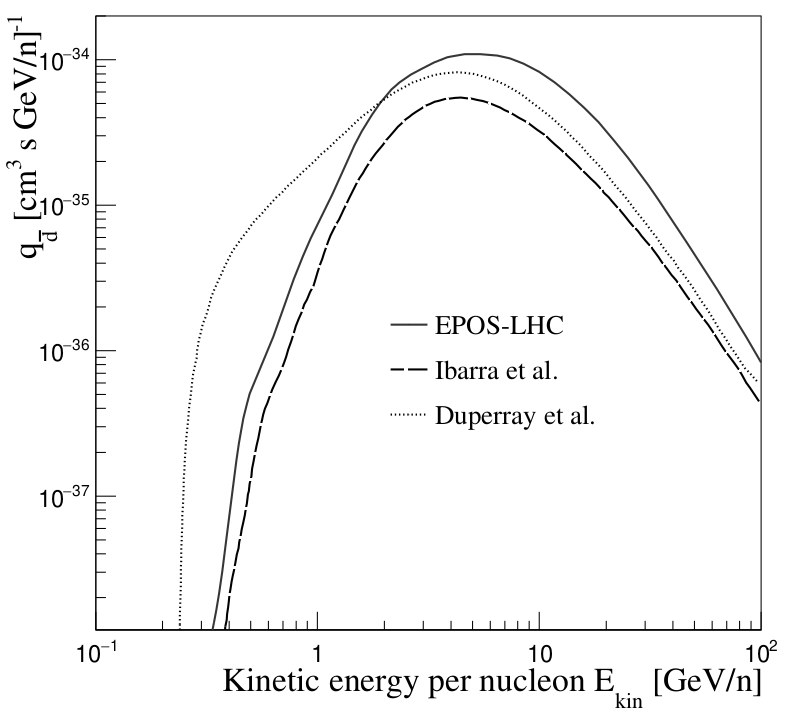}\\
%\includegraphics[height=0.29\textheight]{cap3/qsource_dbar.eps}
%& \includegraphics[height=0.29\textheight]{cap3/qsource_dbar2.eps}\\
\end{tabular}
\caption{Secondary source term for antideuterons as function of kinetic energy per nucleon. On the left side the "Total" contribution from the most important collisions is shown. On the right side, the result is compared to previous works\,\cite{Duperray, Ibarra}.}
\label{fig3}
\end{center}
\end{figure}

\section{Validation}
The first step to calculate the propagation of deuterons and antideuteron in the Galaxy correctly is to be sure the transport model is correctly set up. Special care has to be taken concerning the proton and helium fluxes, since they are the primary incident particles. Additionally, the precise measurement of the Boron-to-Carbon ratio by AMS-02\,\cite{AMS022} helps restricting the diffusion coefficient value, hence, improving the predictability of the model. In Fig.\,4 the fluxes obtained using GALPROP with solar modulation effects taken from\,\cite{Gleeson, Perko}, are compared to AMS-02 data. On the left-top side is the proton flux, where the solid line represents the GALPROP prediction using a Fisk potential of $570$ MeV in the solar modulation model. As a reference, in the same figure the dashed line is the same GALPROP result but using the solar modulation model HelMod\,\cite{Boschini}. In this theoretical approach, the entire CRs transport equation, from the supernova termination shock, to the Earth orbit is Monte Carlo-simulated. On the right-top side of Fig.\,4, the corresponding GALPROP predictions for the Boron-to-Carbon ratio using HelMod, are compared to the AMS-02\,\cite{AMS022} data. In both cases, the relative difference between simulation and data is less than $20\%$, and the solar modulation model used is the force field approximation, again with a Fisk potential of $570$ MeV. As shown at the bottom of this figure, Helium flux is compared as well, the results differ by a maximum of $10\%$ from AMS-02 data.

\begin{figure}[h]
\begin{center}
\begin{tabular}{ll}
\includegraphics[height=0.33\textheight]{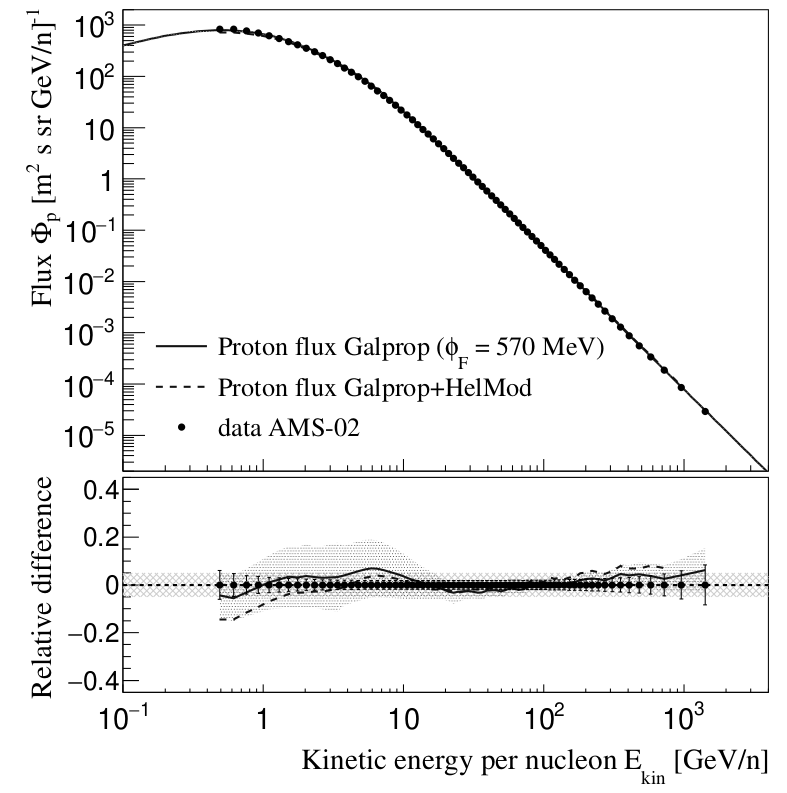}
& \includegraphics[height=0.33\textheight]{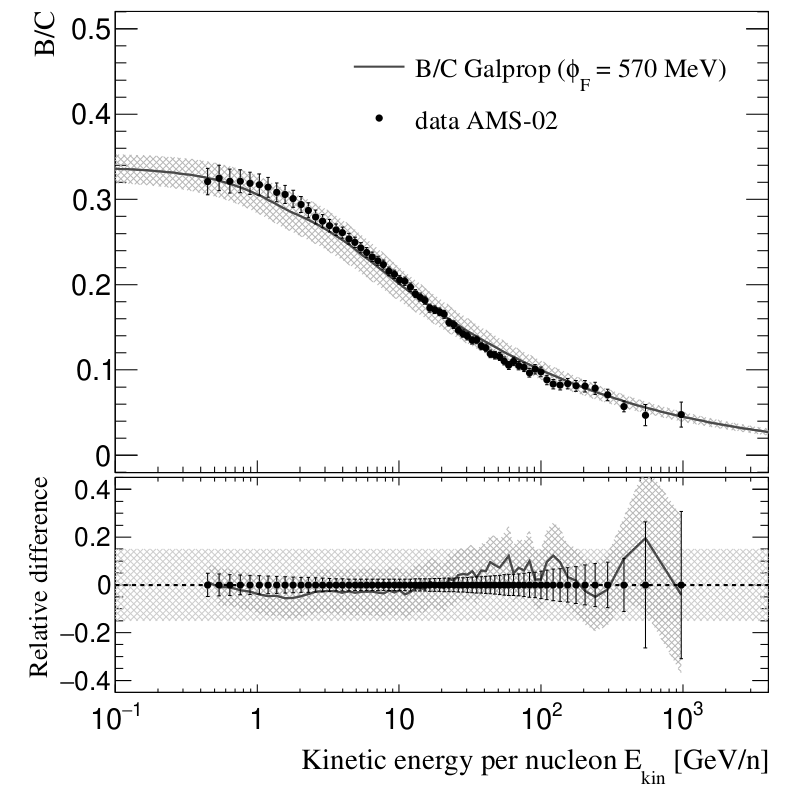}\\
\end{tabular}
\includegraphics[height=0.33\textheight]{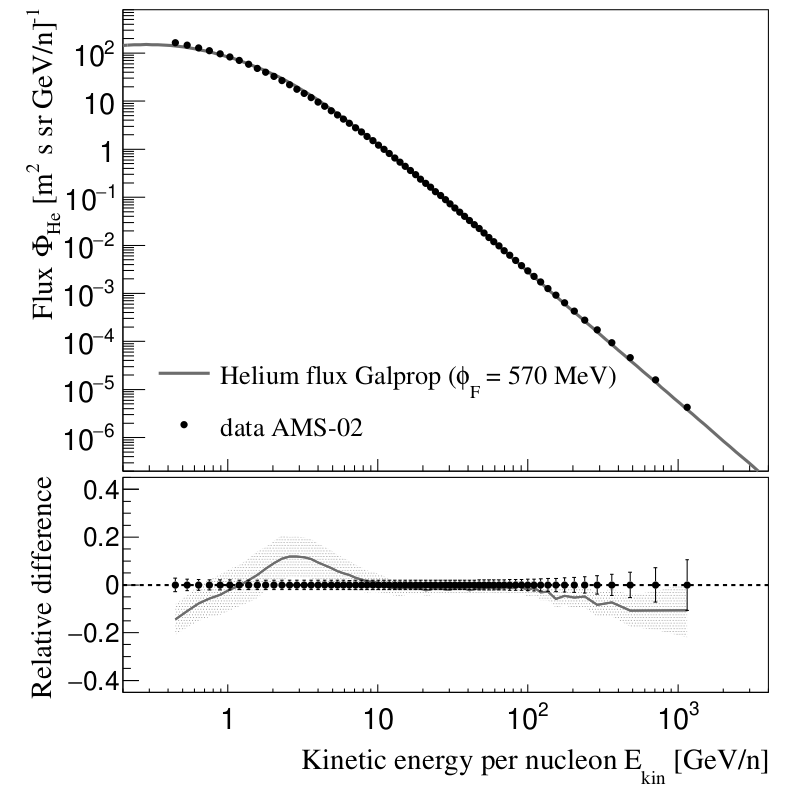}
%\includegraphics[height=0.33\textheight]{cap3/flux_p.eps}
%& \includegraphics[height=0.33\textheight]{cap3/flux_BtoC.eps}\\
%\end{tabular}
%\includegraphics[height=0.33\textheight]{cap3/flux_He.eps}
\caption{Results obtained with GALPROP and the parameters for proton, Boron-to-Carbon ratio and Helium compared to the most recent data from AMS-02\,\cite{AMS022} experiment.}
\label{fig4}
\end{center}
\end{figure}

As already mentioned, in addition to the most abundant CRs species simulated, and validated in Fig.\,4, antiprotons play an essential role in the antideuteron production, and for that reason, their generation in GALPROP has to be analyzed and compared with recent AMS-02\,\cite{AMS022, AMS021} experimental data.

\subsection{Antiprotons}
The next step is to determine the antiproton flux using EPOS-LHC and GALPROP to compare with parametrizations. As in the case of antideuterons, to obtain the antiproton flux, the secondary and tertiary source terms have to be calculated\,\cite{Diego2}. The dominant channels for the production of the secondary antiprotons are: $p+p$ at roughly $50-60\%$ of the total contribution, and $p+He$, $He+p$, and $He+He$ at $10-20\%$ each; while the channels involving heavier incoming CRs contribute only up to a few percent. Thus, here the contributions from the reactions $p+p$, $p+He$, $He+p$, and $He+He$ were included, and antiprotons created in heavier CRs collisions were ignored.
The antiproton multiplicity produced in CRs includes antiprotons from CRs collisions and from hyperon and antineutron decay. This is shown on the left side of Fig.\,5, where the total antiproton source is presented. The distribution shows a maximum at around 2 GeV/n, although ours is slightly broader than previous studies\,\cite{Duperray, Ibarra, Korsmeier}, also shown.

\begin{figure}[h]
\begin{center}
\begin{tabular}{ll}
\includegraphics[height=0.32\textheight]{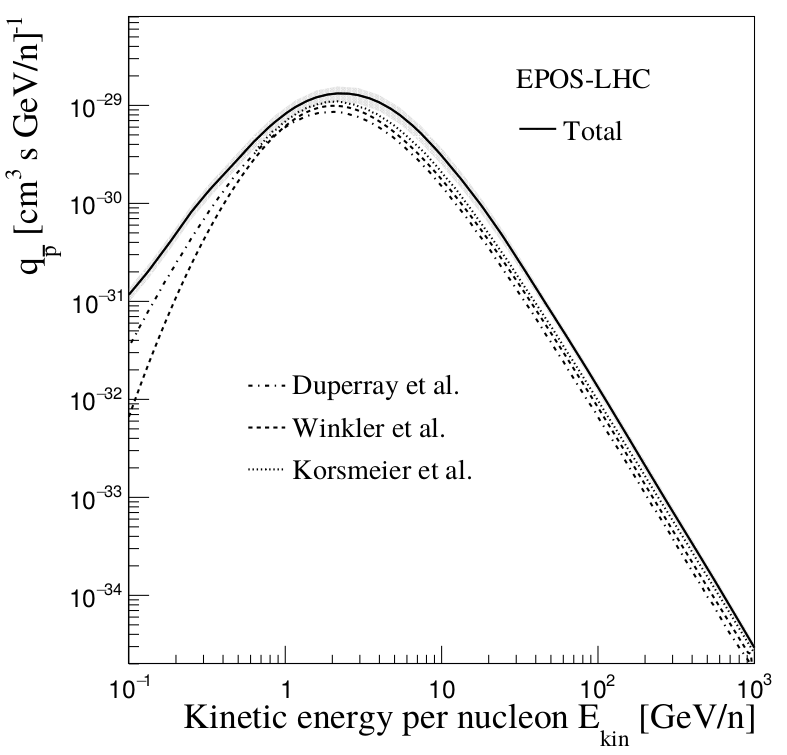}
& \includegraphics[height=0.32\textheight]{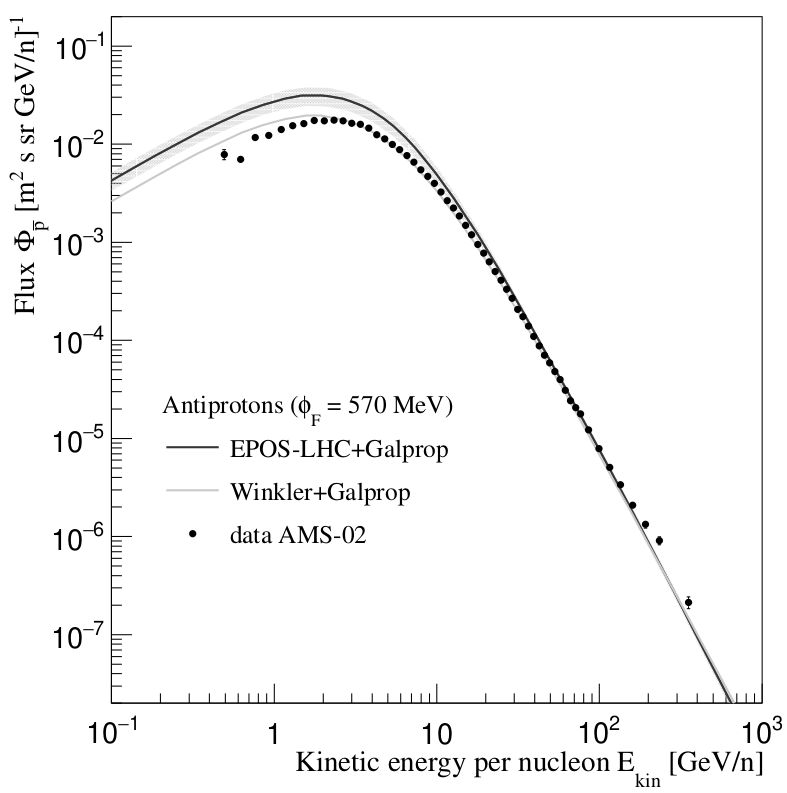} \\
%\includegraphics[height=0.32\textheight]{cap3/qsource_pbar.eps}
%& \includegraphics[height=0.32\textheight]{cap3/flux_pbar2.eps} \\
\end{tabular}
\caption{(Left) Antiproton secondary source term for EPOS-LHC. (Right) Antiproton flux obtained with EPOS-LHC and GALPROP compared to AMS02\,\cite{AMS021} data and Winkler\,\cite{Winkler} prediction.}
\label{fig5}
\end{center}
\end{figure}

Our “Total” antiproton calculation also predicts a larger flux antiproton production than Duperray\,\cite{Duperray}, Korsmeier\,\cite{Korsmeier}, and Winkler\,\cite{Winkler}. An estimate of the source term using EPOS-LHC, but considering the antiproton contribution from antineutron decay as in Korsmeier\,\cite{Korsmeier} shows a lower antiproton production, while conserving the shape of the original EPOS-LHC calculation (Total). This indicates antiproton overproduction generated by EPOS-LHC is related to an excess in the contribution from antineutron decays.
On the right side of Fig.\,5 the antiproton flux is compared to the one obtained by Winkler et al.\,\cite{Winkler} and the data from AMS-02 \,\cite{AMS021}. Our EPOS-LHC simulation predicts a larger antiproton flux below $10$ GeV/n, reaching a magnitude $1.5$ times larger than Winkler calculation. At high energies the antiproton flux calculated with EPOS-LHC shows similar results as the flux from Winkler.

\subsection{Deuterons and Antideuterons}
Once the propagation parameters have been defined and tested, it is safe to proceed with the transport of deuterons and antideuterons, as shown in Fig.\,6. On the left side of the figure, the deuteron flux times $E_{kin}^{2.7}$ is plotted as a function of the deuteron kinetic energy. Besides the deuteron result, the helium flux is included in the same plot to show the good description of data by GALPROP. Data from AMS-1\,\cite{AMS01} and AMS-02\,\cite{AMS021} are presented in the figure together with data from CAPRICE\,\cite{Papini}, which is the only deuteron measurement at high energies. The deuteron flux considering coalescence\,\cite{Diego1} is represented by the thin solid line. The dot-dashed line is the result without coalescence. 

\begin{figure}[h]
\begin{center}
\begin{tabular}{ll}
\includegraphics[height=0.34\textheight]{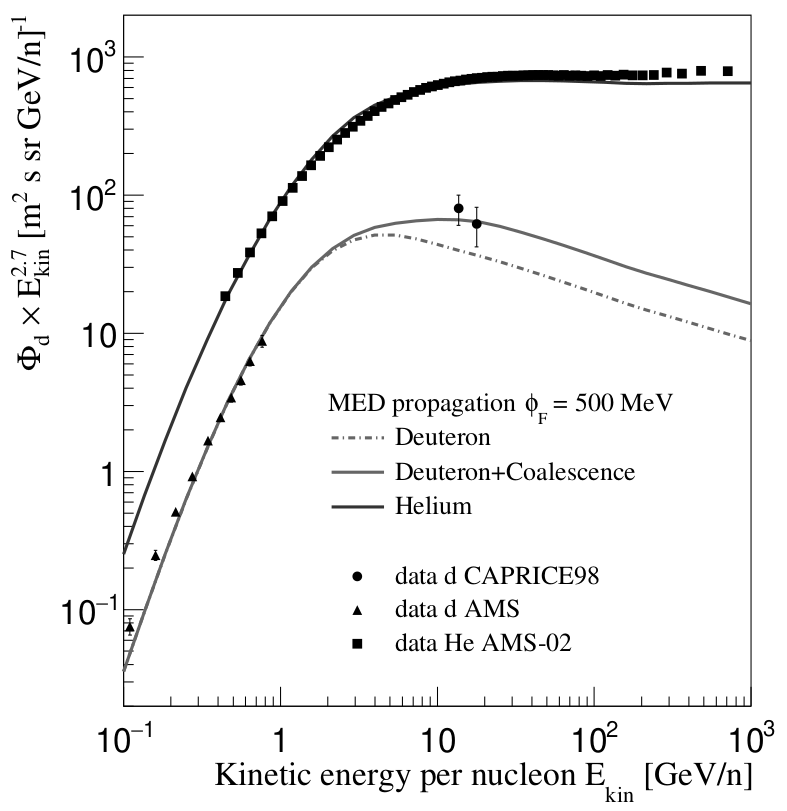}
& \includegraphics[height=0.34\textheight]{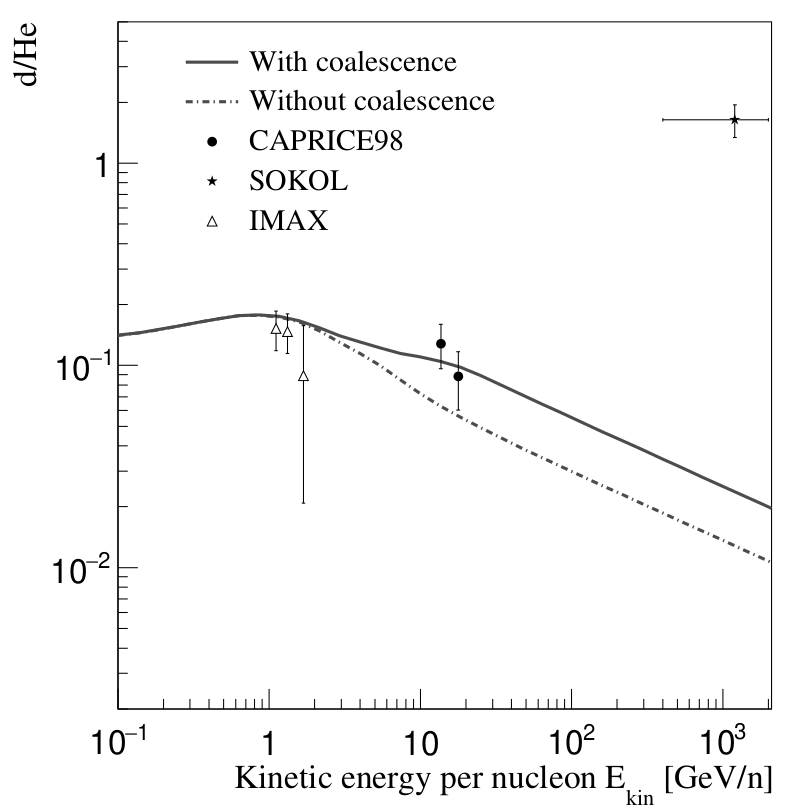}\\
%\includegraphics[height=0.34\textheight]{cap3/flux_d.eps}
%& \includegraphics[height=0.34\textheight]{cap3/dtoHe_ratio.eps}\\
\end{tabular}
\caption{(Left) Deuteron flux produced using QGSJET-II-04 with coalescence (thin solid line) and without coalescence (dot-dashed line) propagated with GALPROP and compared to data from AMS\,\cite{AMS022, AMS01}, CAPRICE\,\cite{Papini}. (Right) Deuteron-to-Helium ratio compared to data from CAPRICE\,\cite{Papini}, SOKOL\,\cite{Turundaevskiy}, and IMAX\,\cite{Nolfo}.}
\label{fig6}
\end{center}
\end{figure}

As can be observed in Fig.\,6, the deuteron flux including coalescence production of deuterons at high energies shows an improved description of data. On the right side of Fig.\,6, the deuteron-to-helium ratio results are presented and compared to data from CAPRICE\,\cite{Papini} and SOKOL\,\cite{Turundaevskiy}. The CAPRICE\,\cite{Papini} ratio was reevaluated with the helium data from AMS-02\,\cite{AMS022}. Again, as in the left side figure, it is observed deuteron simulation containing coalescence production describes data satisfactorily; however, this simulation is not enough to account for the excess predicted by SOKOL\,\cite{Turundaevskiy}.

\section{Result}

The end result of the antideuteron flux calculated with EPOS-LHC as event generator to simulate CR collisions, simultaneously with the afterburner to produce antideuterons through the coalescence model, and the propagation of these particles in the Galaxy by means of GALPROP is presented in Fig.\,7. The first notorious feature on the left side of this figure is that the antideuteron flux evaluated here (solid line) is the largest one compared with previous works. Indeed, the predicted antideuteron flux is around $30\%$ larger than the flux calculated by Duperray et al.\,\cite{Duperray} and nearly to 4 times the one inferred by Ibarra et al.\,\cite{Ibarra}. The last result is not surprising, since from the source term in Fig.\,3 it was evident EPOS-LHC+afterburner produce a higher number of antideuterons as a consequence of the larger cross-section. The predicted antideuteron flux obtained with EPOS-LHC, also shows a wider distribution compared with the mentioned studies, due to a broader high energy contribution originated in a high production cross-section.
An effect of the energy-dependent coalescence momentum and the cross-section derived from that analysis\,\cite{Diego1} is the slight shift of the maximum in the antideuteron flux that now is around $4-5$ GeV/n, when other calculations place it around $3-4$ GeV/n. 

\begin{figure}[!ht]
\begin{center}
\begin{tabular}{ll}
%\includegraphics[height=0.28\textheight]{cap3/fluxdbar.eps}
%& \includegraphics[height=0.28\textheight]{cap3/fluxdbar_DM.eps}\\
\includegraphics[height=0.28\textheight]{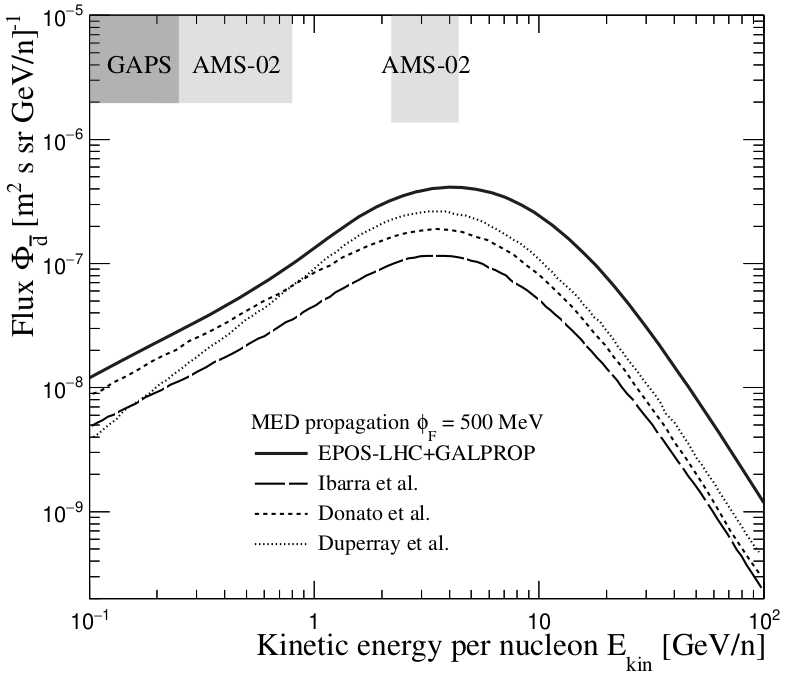}
& \includegraphics[height=0.28\textheight]{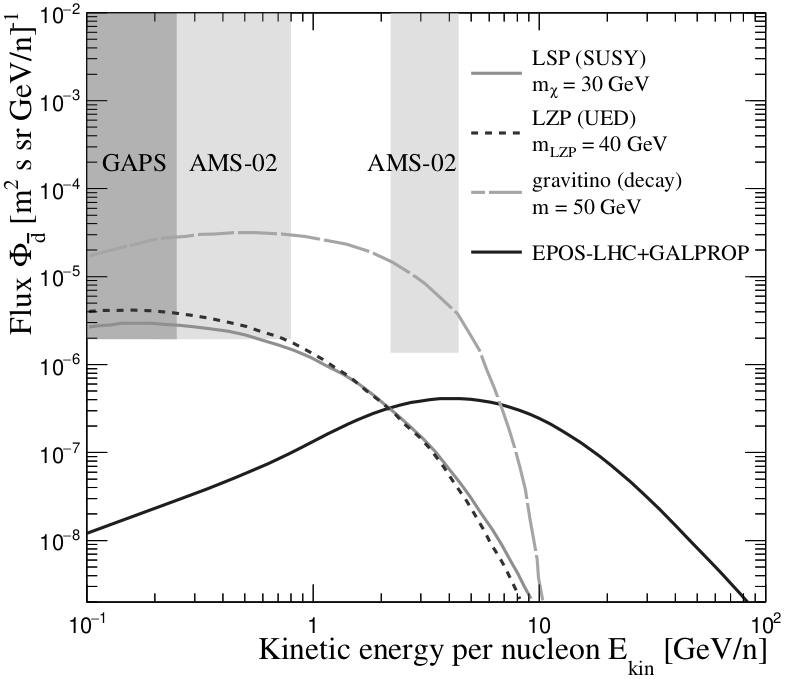}\\
\end{tabular}
\caption{Antideuteron flux at the top of the atmosphere obtained with EPOS-LHC and GALPROP. (Left) The result is compared to previous works from Ibarra\,\cite{Ibarra}, Donato\,\cite{Donato2} and Duperray\,\cite{Duperray}. (Right) The result is compared to antideuteron fluxes expected from dark matter annihilation or decay\,\cite{Donato1, Baer, Dal}. Sensitivity limits for AMS-02 and GAPS are also included.}
\label{fig7}
\end{center}
\end{figure}

If the coalescence momentum would have been considered constant at all energies, then the shape of the distribution would be the same as Duperray et al.\,\cite{Duperray}, or Ibarra et al.\,\cite{Ibarra}. But since the suppression of the coalescence momentum was taken into account near the energy threshold, the low energy region of the flux ($<1 $ GeV) is also suppressed, and therefore the shape of the distribution is modified.
On the right side of Fig.\,7, the resulting antideuteron flux is plotted along with three different expected antideuteron fluxes from dark matter annihilation or decay. These candidates, already shown in Fig.\,1, continue having a considerable larger flux below $1$ GeV/n compared to the secondary, despite its increase using EPOS-LHC.

\section{Conclusion}

In conclusion, the antideuteron flux obtained here shows a larger magnitude compared to other studies, and a slight difference in the shape distribution, as a consequence of the energy dependence of the coalescence parameter (Fig.\,7). Although the resulting flux is $1.3$ times larger than the predicted by Duperray et al.\,\cite{Duperray}, and four times than Ibarra et al.\,\cite{Ibarra}, it is still below the expected sensitivities of AMS-02 and GAPS. These findings strengthen the idea that if these experiments detect cosmic antideuterons, they are probably from exotic sources.

\section{Acknowledgements}
The authors would like to thank J. Martinez Mendoza, E. Lopez Pineda and S. Aguilar Salazar for technical aid, as well as CONACyT and PAPIIT-DGAPA: IN109617 for financial support.

\end{document}